\documentstyle[preprint,aps,epsfig]{revtex}
\begin{document}
\tightenlines
\draft
\title{The Strange Quark Mass From Finite Energy Sum Rules}
\author{Kim Maltman} 
\address{Department of Mathematics and Statistics, York University, \\
          4700 Keele St., Toronto, Ontario, CANADA M3J 1P3 \\ and}
\address{Special Research Centre for the Subatomic Structure of Matter, \\
          University of Adelaide, Australia 5005}
\maketitle
\begin{abstract}
The strange quark mass is determined from a study
of the correlator $\langle 0\vert T\left( J_s J^\dagger_s\right)\vert 0
\rangle$ (where $J_s\equiv\partial_\mu \bar{s}\gamma^\mu u$,
the divergence of the strange vector
current) using a family
of finite energy sum rules recently shown to be very accurately
satisfied in the isovector vector channel.  
It is shown that the match between the OPE and hadronic representations
for the sum rules employed is extremely good
once one fixes the overall scale,
$\left( m_s-m_u\right)^2$, which normalizes the OPE side.  
A value $m_s(4\ {\rm GeV}^2)=115\pm 8\ {\rm MeV}$,
corresponding to $m_s(1\ {\rm GeV}^2)=159\pm 11\ {\rm MeV}$, is
obtained, with theoretical and experimental uncertainties
contributing roughly equally to the combined error.
\end{abstract}
\section{Introduction}
The light quark masses, despite being fundamental parameters of
the Standard Model, are surprisingly poorly known at present.
Since, in the standard (``large condensate'')
scenario for the chiral counting in Chiral Perturbation Theory (ChPT), 
however, the ratio
$m_s/(m_u+m_d)=24.4\pm 1.5$ is known\cite{leutwylerqmasses}, fixing
either $m_s$ or $m_u+m_d$ is sufficient to determine
the other.  In this paper we focus on $m_s$, which
has been the subject of several recent investigations,
employing both
QCD sum rules\cite{jm,narison95,cps,cfnp,jamin,km3388,dps,kmtau,pp98,cdh98} 
and lattice QCD
\cite{rt97,al97,alrev,gough97,sesam,cppacs,ape,qcdsf,k98,gimenez,jlqcd,bsw99}.

On the lattice, reliable results are presently restricted to the
quenched approximation and vary somewhat depending on the fermion
implementation and on whether one chooses $m_K$ or $m_\phi$ when 
matching to physical meson masses.
Unquenched calculations are at a more preliminary
stage, but suggest unquenching is likely
to lower quenched results\cite{rt97,sesam,cppacs,k98}.
Recent quenched results for $m_s$ (where here, as in what
follows, all masses, unless otherwise specified, will be
quoted at a scale $\mu=2$ GeV in the $\overline{MS}$ scheme) are
$110\pm 20 \pm 11$ MeV\cite{rt97},
$122\pm 20$ MeV\cite{al97}, $95\pm 16$ MeV\cite{gough97},
$115\pm 2\ (143\pm 6)$ MeV\cite{cppacs} (for $m_K\ (m_\phi )$
input), $111\pm 12$ MeV\cite{ape}, $98.1\pm 2.4$ MeV\cite{qcdsf}
(update as quoted by Kenway\cite{k98}), $130\pm 2\pm 18$ 
MeV\cite{gimenez}, $129\pm 12$ MeV\cite{jlqcd} and
$95\pm 26$ MeV\cite{bsw99}.  
For dynamical
($N_f=2$) simulations, the CP-PACS results $m_s\sim 70\ (80)$
MeV (with $m_K \ (m_\phi )$ as input) and SESAM result $151\pm 30$
MeV\cite{sesam} do not appear compatible. (Note, however, that
there are unresolved issues of chiral extrapolations, renormalization
constants, and the lack of extrapolation to the continuum limit in
the SESAM result; for a further discussion of these issues see
Section
6 of the review by Bhattacharya and Gupta\cite{rt97}, and 
Allton\cite{alrev}.)

Recent attempts to extract $m_s$ using QCD sum rules are
of four types.  Two employ flavor-breaking differences
of vector and/or axial vector
current correlators, either the isovector-hypercharge
difference\cite{narison95,km3388} or the isovector-strange
difference\cite{kmtau,pp98,cdh98}.
The others involve analyses of the strange 
scalar\cite{jm,cps,cfnp,jamin} and strange pseudoscalar\cite{jm,dps} channels.

For the vector current isovector-hypercharge difference, isospin-breaking
corrections are crucial due to
the high degree of cancellation on the hadronic side of the sum rule.
These are, for natural
reasons\cite{km3388,kmcvc}, larger than one would naively 
expect, and can only
be obtained indirectly\cite{km3388}.  The high degree of cancellation
also magnifies experimental and theoretical uncertainties,
leading to significant errors
on $m_s$ ($m_s=107\pm\sim 40$ MeV\cite{mw98}, if
one takes the isovector 
spectral function from 
hadronic $\tau$ decay data\cite{ALEPH,OPAL}).  

The isovector-strange
difference, which employs $\tau$ decay data for the isovector
and strange spectral functions, is subject to complications
resulting from (1) very poor convergence of the
perturbative series for the longitudinal
contributions on the OPE side\cite{kmtau,pp98},
and (2) strong cancellation, 
which again significantly magnifies
the effect of the (otherwise rather small) experimental errors.
Combining errors
in quadrature and translating to the scale $2$ GeV,
one finds 
$m_s=163\pm 50$ MeV\cite{kmtau},
$143\pm 42$ MeV\cite{pp98} and $157\pm 37$ MeV\cite{cdh98}.  

The third approach involves
the correlator
$\langle 0\vert T\left( J_{A,s} J^\dagger_{A,s}\right)\vert 0
\rangle$ 
(where $J_{A,s}\equiv\partial_\mu \bar{s}\gamma_5\gamma^\mu u$)\cite{jm,dps}.
In this case, the
$K$ pole contribution to $\rho (s)$ is known,
but contributions from the continuum ($K(1460)$, $K(1830)$ resonance
region) are not.
Both analyses of this type employ a sum-of-Breit-Wigner modulation
of the near-threshold tree-level ChPT form to deal with this problem,
together with some assumptions about the relative
size of the two unknown resonance decay constants.  As discussed in
Refs.~\cite{cfnp,bgm}, this type of normalization procedure
for a continuum spectral ansatz can involve
potentially significant systematic errors.
For example, in the scalar strange
channel, it leads to a rather poor
representation of the actual spectral function (compare
the model versions of Refs.~\cite{jm,cps} with that
of Ref.~\cite{cfnp}).  
In addition,
because both analyses employ Borel transformed (SVZ) sum rules\cite{svz}, 
they involve the conventional
``continuum'' (or local duality) approximation
to $\rho (s)$ above $s=s_0$ (where the
``continuum threshold'', $s_0$, is a parameter of the analysis).
Since, for Borel masses, $M$, in the stability windows of
both analyses, $M^2<<s_0$ is far from being
satisfied, there are potential systematic errors
resulting from violations of local duality in the 
``continuum'' region.  (That contributions from the ``continuum''
region are not numerically negligible can also be inferred, e.g.,
from the figures of Ref.~\cite{dps}).  It is worth noting that 
violations of local duality are, in general, expected in kinematic
regimes where typical resonance widths are comparable 
to or smaller than typical resonance separations.  This has been
quantified in
the isovector vector channel, where the hadronic spectral function
is known experimentally; violations
of local duality in that channel, even at scales $\sim m_\tau^2$,
are seen to be, in general, rather large\cite{kmfesr}.
Taking the results from Refs.~\cite{jm,cps}
which employ $\Lambda_{QCD}=380$ MeV
(which best corresponds to recent
experimental values for $\alpha_s(m_\tau^2)$\cite{ALEPH,OPAL}), 
one finds, for $m_s$, the values $\simeq 100$ MeV\cite{jm}
(see Figure 5) and $115\pm 19$\cite{dps} (where the errors do not reflect
the systematic uncertainties discussed above).

The last approach involves the correlator, relevant to the strange
scalar channel,
\begin{eqnarray}
\Pi (q^2)\ &=&\ i\int\, dx\, e^{iq\cdot x}\langle 0\vert
T\left( J_s(x)\, J_s^\dagger (0)\right)\vert 0\rangle \nonumber\\
&=&  i \left( m_s-m_u\right)^2\, 
\int\, dx\, e^{iq\cdot x}\langle 0\vert
T\left( \bar{s}u(x)\, \bar{u}s(0)
\right)\vert 0\rangle\ ,
\label{scalarcorrelator}
\end{eqnarray}
which has been analyzed, in Refs.~\cite{jm,cps,cfnp,jamin},
using the Borel transformed sum rule method.  This channel
is a favorable one for sum rule analyses because
the hadronic spectral function, $\rho (s)$, turns out to be
rather well-determinable
(albeit indirectly) from experimental data.
The reason is that since, experimentally, 
$K\pi$ scattering is known to be purely elastic
up to $s=(1.58\ {\rm GeV})^2$\cite{LASS,dunwoodie},
the phase of the timelike scalar
$K\pi$ form factor, $d(s)$, is known up to this point.
$d(s)$, moreover, satisfies an Omnes
representation
\begin{equation}
d(s)=d(0)p(s)\, exp\left[ {\frac{s}{\pi}}\int_{th}^\infty\, dt
{\frac{\delta (t)}{t(t-s-i\epsilon )}}\right]\ ,
\label{omnes}
\end{equation}
where $p(s)$ is a polynomial normalized to $1$ at $s=0$,
$\delta (s)$ is the phase of $d(s)$, and the
normalization $d(0)$ is known from ChPT and $K_{e3}$\cite{glmff}.  In order to
satisfy quark counting rules for the asymptotic behavior of
$d(s)$, $\delta$ must $\rightarrow (N+1)\pi$ as $s\rightarrow\infty$
if $p(s)$ is of degree $N$.  In previous analyses $p(s)$ has 
been assumed to be $1$ (or, equivalently, 
very slowly varying in the range of $s$ important to the analysis).
The fact that, at the edge of the experimentally explored region,
$s=(1.7\ {\rm GeV})^2$,
the $I=1/2$ $K\pi$ phase shift is very nearly
$\pi$\cite{LASS,dunwoodie,estabrooks}, the asymptotic
value required if $p(s)\equiv 1$, suggests, first,
that this assumption is rather
plausible and, second, that the treatment of the
high energy behavior of the phase, $\delta (s)$,
employed in Ref.~\cite{cfnp}, {\it i.e.} 
$\delta (s)=\pi$ for $s>(1.7\ {\rm GeV})^2$, is sensible.
We will also see below that, making these assumptions,
one obtains extremely well-satisfied sum rules, providing
strong {\it post facto} justification as well.  With these assumptions,
the $K\pi$ contribution to $\rho$ is completely
determined, and multiparticle contributions 
begin only above $s\sim 2.5\ {\rm GeV}^2$. 
The resulting $K\pi$ contribution to $\rho (s)$ is, moreover,
much smaller in the 
$K_0^*(1950)$ than in the $K_0^*(1430)$ region
(see Fig. 2 of Ref.~\cite{cfnp}).  Since
the $K_0^*(1950)$ $K\pi$ branching fraction
is known to
be $\sim 50\%$\cite{pdg98}, the remaining multiparticle contributions
to $\rho$, in the $K_0^*(1950)$ region, will also be small.
This means that, to the extent that
one may expect
multiparticle contributions to $\rho$
to be resonance dominated, (1) such contributions will play only 
a small role in the full spectral
function out to $s\sim 4\ {\rm GeV}^2$, and (2) in this region,
the magnitude of these contributions can
be obtained by combining information from
the Omnes-generated
$K\pi$ component and the known
experimental $K_0^*(1950)$ $K\pi$ branching fraction.
We do not quote results 
from the analyses of Refs.~\cite{jm,cps}
since the spectral ansatz employed 
there has been shown to be incompatible with
the Omnes representation\cite{cfnp}.  
Results from Refs.~\cite{cfnp,jamin} are $m_s=\, 91\rightarrow 116\ 
{\rm MeV}$\cite{cfnp} and $116\pm 22\ {\rm MeV}$\cite{jamin}.
The main potential uncertainty in these analyses
is that associated with violations of local duality
in the ``continuum'' region, $s>s_0$. 
That continuum contributions are non-negligible
may be seen from the lack of a stability plateau for $m_s$ in
the analysis of Ref.~\cite{jamin}.
Finite energy sum rules constructed using the extracted
values of $m_s$ from Refs.~\cite{cfnp,jamin} and
the Omnes-generated spectral ansatz of Ref.~\cite{cfnp} 
(also used in Ref.~\cite{jamin}) as input have been shown to be
moderately well satisfied\cite{kmfesr}.  The residual deficiencies
could arise from either shortcomings in the spectral ansatz or 
approximations on the OPE side of the sum rules.  
We will investigate this question
in what follows.

\section{The Strange Scalar Channel Analysis}

Unitarity and analyticity ensure that correlators like
$\Pi (q^2)$, defined above,
satisfy either appropriately subtracted conventional
dispersion relations or finite energy sum rules (FESR's).
Borel transformation of the former leads to the conventional\cite{svz}
(SVZ) form of QCD sum rules.  In what follows we choose to
employ FESR's,
which are based on Cauchy's
theorem, rather than the Cauchy representation theorem,
and work with the
``Pac-man'' contour, which traverses both sides of the physical
cut, and is closed by a circle of radius $s_0$ in the complex-$s$
plane.  One then has
\begin{equation}
\int_{s_{th}}^{s_0}\, ds\, \rho (s) w(s)\, =\,
{\frac{-1}{2\pi i}}\, \oint_{\vert s\vert =s_0}\, ds\, w(s)\Pi (s)\ ,
\label{basicfesr}
\end{equation}
valid for any function $w(s)$ analytic in the region of the contour,
where $\rho (s)$ is the spectral function and
$s_{th}$ the relevant physical threshold.  To make this relation
useful, one wishes to use the physical spectral function, whose
extraction was described in the Introduction, on the LHS, and the
OPE for the correlator on the RHS. Since the OPE is proportional
to $\left( m_s-m_u\right)^2$, this provides a method for 
extracting $m_s-m_u$.  

To make the results of this analysis reliable,
$s_0$ should satisfy two criteria: (1) it should be
large enough to make the OPE representation of the RHS accurate,
and (2) it should be small enough that one still has 
experimental information on $\rho (s)$ for $s<s_0$ on the LHS.
The latter criterion, in the present case, restricts us to
the region up to and including the $K_0^*(1950)$
(i.e., $s_0<4\ {\rm GeV}^2$ or so).  Since,
even at $s=4\ {\rm GeV}^2$, resonance widths and
separations are comparable, one expects local duality
to be violated, creating potentially significant uncertainties
on the OPE side.  It has been argued,
however, that even if local duality is not exactly
satisfied for given timelike $s_0$, the OPE may, nonetheless, provide
a good representation of the correlator on most of
the circle $\vert s\vert =s_0$ (i.e., except possibly for a region of
hadronic scale near the timelike real axis)\cite{pqw}.  
The isovector vector channel,
for which the spectral function has been extracted
from hadronic $\tau$ decay data,
provides explicit empirical confirmation of this argument.  First,
the conventional FESR treatment of these decays\cite{taudecay}
is very well satisfied,
presumably because it involves
a weight function (determined by kinematics) with a double
zero at $s=s_0$ which cuts out contributions from the
potentially dangerous part of the contour.
Second, one may demonstrate that using weights $s^k$ (with
no such zero at $s=s_0$) leads to significant violation of
local duality, even at scale $s_0=m_\tau^2$, whereas
FESR's based on weights having either a single or double zero
at $s=s_0$ are all extremely well satisfied, even at
scales significantly below $s_0=m_\tau^2$\cite{kmfesr}.
Using such FESR's one may, thus, hope to construct sum rules in which
the use of the ``continuum'' ansatz for the
high energy portion of $\rho (s)$ may be avoided.

In what follows, therefore, we 
study FESR's for the correlator, $\Pi (q^2)$, 
using weights of the form
\begin{equation}
w_A(s)=\left( 1-{\frac{s}{s_0}}\right)\left( A-{\frac{s}{s_0}}\right)\ .
\label{NORM}
\end{equation}
In Eq.~(\ref{NORM}), $A$ is a free parameter, to be chosen in such a way
as to optimize the reliability
of the analysis.  
Since (1) instanton effects are known exactly only in dilute gas
single-instanton-background approximation\cite{bbb93,nason} and (2)
the instanton density used to obtain these results is known to be
altered by quark and gluon
condensate effects\cite{svzinst}, the magnitude of the resulting
instanton contributions (as stressed by the authors
of Refs.~\cite{bbb93,nason})
can be considered only a rough guide.  
For this reason, we restrict our attention to
weights which, after
integration, yield 
instanton effects (including corrections to the coefficients of both
the unit
operator and quark condensate in the OPE, as given in Ref.~\cite{nason})
which are small for all $s_0$ employed.  While any $A>1$
turns out to satisfy this criterion, $A=2$ creates an optimal suppression
and we will display results below for this value of $A$.

We now discuss the input on the hadronic side of the
sum rules.  The $K\pi$
and multiparticle portions of $\rho (s)$
are obtained, using the Omnes representation, 
experimental $K\pi$ phases, and the experimental $K\pi$ branching
fractions of the $K_0^*(1430)$ and $K_0^*(1950)$, as described
in the Introduction.  We employ two slightly different implementations
of this procedure, the difference in the resulting $m_s$ values
serving to quantify the effect of uncertainties in our knowledge of the
hadronic spectral function. In the first implementation,
the fit to the experimental $K\pi$ phases obtained in Ref.~\cite{jm} is
used as input to the Omnes representation, and
multiparticle contributions are allowed in both the
$K_0^*(1430)$ and $K_0^*(1950)$ regions, constrained by the
Particle Data Group's $K\pi$
branching fractions\cite{pdg98}. In the
second implementation we employ the LASS fit\cite{LASS,dunwoodie} 
to the $K\pi$ phase shifts and inelasticities.  For this fit,
the form of the parametrization of phase shift data used is slightly
different from that of Ref.~\cite{jm}.  In addition, since
the fit to the $K\pi$ amplitude is 
still purely elastic in the $K_0^*(1430)$
region, in employing this implementation
we allow multiparticle contributions to the spectral
function only from the $K_0^*(1950)$ region. (Note that the background 
effective range parameters, $a$, $b$, and
$K_0^*(1430)$ resonance parameters, $m$, $\Gamma$,
were mis-recorded in Ref.~\cite{LASS},
and should actually be $m=1.412\ {\rm GeV}$,
$\Gamma =.294\ {\rm GeV}$\cite{dunwoodie,gnrmp},
and $a=2.19\ {\rm GeV}^{-1}$,
$b=3.74\ {\rm GeV}^{-1}$\cite{dunwoodie}.  It has been checked that
these provide a satisfactory
representation of the amplitude and phase shift data of Ref.~\cite{LASS}.)

On the OPE side, it is convenient to work with $d^2\Pi (Q^2)/\left(
dQ^2\right)^2\equiv \Pi^{\prime\prime}(Q^2)$, 
which satisfies a homogenous RG equation, so
all scale dependence can be absorbed into the running mass
and coupling.  To improve the convergence of the perturbative
series, and significantly reduce residual scale dependence, we
follow standard practice and evaluate the OPE side
using the ``contour improvement'' 
prescription of Ref.~\cite{contourimproved}, 
first summing logarithms through the
renormalization scale choice $\mu^2=Q^2$, and then
performing the integral around 
the circle $\vert s\vert =s_0$ numerically,
using the exact solutions for the
running mass and coupling which follow from the four-loop-truncated $\beta$
and $\gamma$ functions\cite{beta4,gamma4}.

The expressions for the dimension $D=0,2,4,6$
contributions to $\Pi^{\prime\prime}(Q^2)$ 
are all known from previous 
work\cite{jm,cps,cfnp} and, for $\mu^2=Q^2$, are:
(1) for $D=0$\cite{jm,cps},
\begin{eqnarray}
\left[ \Pi^{\prime\prime}(Q^2)\right]_{D=0}&=& 
\frac{3(m_s-m_u)^2(Q^2)}{8\pi^2 Q^2}\left\{
1 + \frac{11a(Q^2)}{3} + {a(Q^2)^2}\left(
\frac{5071}{144}-\frac{35}{2}\zeta(3)\right)\right.\nonumber\\
& &\left.\qquad +{a(Q^2)^3}\left(\frac{1995097}{5184}-\frac{\pi^4}{36}
-\frac{65869}{216}\zeta (3)+\frac{715}{12}\zeta (5)\right)\right\}\ ,
\label{d0}
\end{eqnarray}
where $a(Q^2)=\alpha_s (Q^2)/\pi$ and $\zeta (N)$ is the Riemann
zeta function; (2) for $D=2$\cite{jm,cps},
\begin{equation}
\left[ \Pi^{\prime\prime}(Q^2)\right]_{D=2}= 
-\frac{3(m_s-m_u)^2(Q^2)m_s^2(Q^2)}{4\pi^2 Q^4}\left\{ 1 +
\frac{28a(Q^2)}{3}
+{a(Q^2)^2}\left(\frac{8557}{72}-\frac{77}{3}\zeta(3)\right)\right\}\ ;
\label{d2}
\end{equation}
(3) for $D=4$\cite{jm,cps},
\begin{eqnarray}
\left[ \Pi^{\prime\prime}(Q^2)\right]_{D=4} &=& 
\frac{(m_s-m_u)^2(Q^2)}{Q^6}\left\{ 2\langle m_s\bar uu\rangle
\left(1 + \frac{23}{3}{a(Q^2)}\right)
-\frac{1}{9} I_G\left( 1 + \frac{121}{18} {a(Q^2)}
\right)
\right.\nonumber\\
& &\left. \ \ 
+ I_s\left( 1+\frac{64}{9}{a(Q^2)}\right) 
-\frac{3}{7\pi^2}m_s^4(Q^2)\left(\frac{1}{a(q^2)}+
\frac{155}{24}\right)\right\}\ ,
\label{d4}
\end{eqnarray}
where $I_G$ and $I_s$ are the RG-invariant $D=4$ condensate combinations
of Ref.~\cite{cps}; and (4) for $D=6$\cite{jm},
\begin{equation}
\left[ \Pi^{\prime\prime}(Q^2)\right]_{D=6} =
\frac{(m_s-m_u)^2(Q^2)}{Q^8}\left\{ 3 m_s\langle
g\bar{u}\sigma F u\rangle -\frac{32}{9}\pi^2 \rho_{VSA} a
\left( \langle \bar{s}s\rangle^2 + \langle \bar{u}u\rangle^2
+9  \langle \bar{s}s\rangle \langle \bar{u}u\rangle\right)\right\}\ ,
\label{d6} 
\end{equation}
where $\rho_{VSA}$ is a parameter describing the deviation
of the four-quark condensates from their 
vacuum saturation values.
In the above equations, factors of $m_{u,d}$
have been dropped, except in the overall
normalization $(m_s-m_u)^2$.

For the input required on the OPE side of the sum rule we employ
the following:  
$\alpha_s(m_\tau^2)=.334\pm .022$\cite{ALEPH}
(needed as initial condition for
the RG running of $\alpha_s$); for the $D=4$
condensates, 
$\langle \alpha_s G^2\rangle = 0.07\pm .01\ {\rm GeV}^4$\cite{narison97},
$\left( m_u+m_d\right)\langle \bar{u}u\rangle =-f_\pi^2 m_\pi^2$, and
$0.7< r_c \equiv
\langle \bar{s}s\rangle /\langle \bar{u} u\rangle <1$\cite{jm,cps};
for the $D=6$ condensates, $\langle g\bar{q}\sigma Fq\rangle
=\left( 0.8\pm 0.2\ {\rm GeV}^2\right)\langle \bar{q} q\rangle$\cite{jm}
and $\rho_{VSA}=5\pm 5$ (i.e., allowing, to be conservative,
up to an order of magnitude 
deviation from vacuum saturation). We evaluate
$m_s-m_u$ by optimizing the match of the OPE and hadronic sides of
the sum rules in question
in the range $3\ {\rm GeV}^2\leq s_0\leq 4\ {\rm GeV}^2$,
for which the dominant $D=0$ contour-improved series is well behaved.
To convert the extracted values
of $m_s-m_u$ to those for $m_s$, we take the central ChPT value
for $m_s/m_u$\cite{leutwylerqmasses}.

The errors on the extracted value of $m_s$ associated with those on
the input OPE parameters, together with their sources, are
as follows (where we list only those cases where the error is greater
than $0.1$ MeV):  due to $\alpha_s(m_\tau^2)$,
$\pm 1.0$ MeV; due to
$\rho_{VSA}$, $\pm 0.7$ MeV; due to $\langle \alpha_s G^2\rangle$, 
$\pm 0.2$ MeV;
due to $r_c$, $\pm 0.2$ MeV.  More significant, on the OPE side, is
the error associated with the truncation of the $D=0$ contributions
at four-loop order.  If we assume continued geometric growth of the
expansion coefficients, including this estimate for the next order
term lowers $m_s$ by $2.8$ MeV.  In the final results below we
have included this estimate of the five-loop order term
in arriving at the quoted value, and assigned
an error of {\it twice} the difference between the four-loop
result and five-loop estimate.  If added in quadrature, this source
of error swamps all others on the OPE side.

The sources and errors on $m_s$
associated with uncertainties on the hadronic side
of the sum rules (where greater than $0.1$ MeV) are as follows:
due to the error on the $K_0^*(1430)$ $K\pi$ branching
fraction\cite{pdg98} (in the case of the first implementation described
above, where multiparticle contributions are allowed in the
$K_0^*(1430)$ region), $\pm 3.5$ MeV;
due to that on $m_{K_0^*(1430)}$, $\pm 0.6$ MeV; due to 
that on $\Gamma_{K_0^*(1430)}$,
$\pm 3.8$ MeV; and due to that on $\Gamma_{K_0^*(1950)}$, $\pm 0.4$ MeV.  There
is also a $0.4$ MeV difference between the extracted central values
for the two different implementations of the hadronic spectral
ansatz; this
uncertainty is, however, swamped by that associated
with the treatment of multiparticle contributions in the
$K_0^*(1430)$ region in the case of the
first implementation.  We treat the latter as a systematic error
common to both hadronic implementations,
even though it is, strictly speaking,
not consistent to include
it in the case of the second
implementation, based on the LASS fit, which is purely elastic
in the $K_0^*(1430)$ region.

Taking the average of the central values associated with the two different
hadronic implementations, and adding errors in quadrature, we then
arrive at our final result,
\begin{equation}
m_s\left( 4 \ {\rm GeV}^2\right)=\left( 115.1\pm 5.7 \pm 5.2\right) 
\ {\rm MeV}\ ,
\label{final}
\end{equation}
where the first error is associated with the OPE 
side of the sum rule (and is completely
dominated by our estimate of the error due to trunctation of the
$D=0$ series) and the second with uncertainties on
the hadronic side (where
errors associated with the inelasticity in the $K_0^*(1430)$ region
and with the uncertainty in the $K_0^*(1430)$ width are roughly
equally important and, again, dominate all other sources).
This result corresponds to
\begin{equation}
m_s\left( 1 \ {\rm GeV}^2\right)=\left( 158.6\pm 7.9 \pm 7.2\right) 
\ {\rm MeV}\ .
\label{onegev2}
\end{equation}

In Figure 1 we display the match between the OPE and hadronic
sides of the $A=2$ sum rule in the fit window, 
for the optimized value of $m_s$.  The agreement is
obviously excellent, increasing our confidence in both the
assumptions about the form of the Omnes representation of $d(s)$,
and the value of $m_s$ that results.

\section{Summary}

We have determined $m_s$ from a FESR analysis in which the
hadronic side of the sum rules employed is well determined by
experimental data, modulo questions about the presence or
absence of a polynomial factor in the Omnes representation
of $d(s)$ and the high energy behavior of the phase
$\delta (s)$.  The excellent agreement between OPE and hadronic
sides shown in the Figure argues strongly, albeit {\it post facto},
for the reliability of the input assumptions $p(s)=1$ and $\delta (s)=\pi$
for $s>(1.7\ {\rm GeV})^2$.  We have
attempted to make conservative estimates of the errors
associated with uncertainties on the hadronic and OPE sides
of the sum rules and conclude that $m_s\left( 4 \ {\rm GeV}^2\right)
=115\pm 8\ {\rm MeV}$, compatible with results from recent
quenched lattice calculations.  
It is worth noting that systematic uncertainties 
in the Borel transformed sum rule approach
associated
with the use of the ``continuum ansatz'' for the high energy
portion of the hadronic spectral function
are potentially significant in size on
the scale of the errors quoted above.  In Ref.~\cite{cps}, for
example, a variation of $\pm 9$ MeV in $m_s\left( 1\ {\rm GeV}^2\right)$
results from variations of $\pm 0.5\ {\rm GeV}^2$ in
the continuum threshold
parameter, $s_0$.  Such uncertainties are, of course, 
absent in the FESR approach, and help to keep the final errors small.

\vskip .5in\noindent
 \begin{figure} [htb]
\centering{\
\psfig{figure=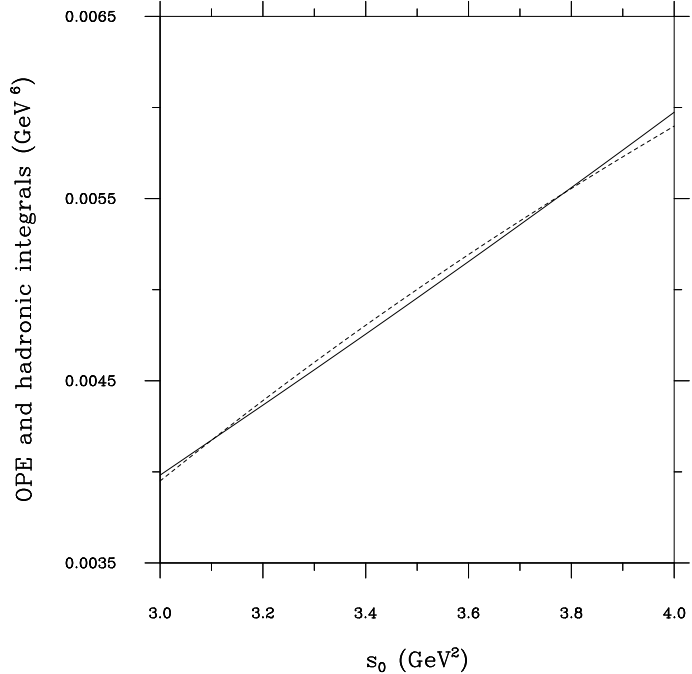,height=10.cm}}
\vskip .3in         
\caption{The OPE ($I_{OPE}$) and hadronic 
($I_{had}$) sides of the $A=2$ FESR defined in the
text.  The solid line gives the OPE side, the dashed
line the hadronic side.}
\label{figone} 
\end{figure}
 
\acknowledgements
The author would like to thank Rajan Gupta, Tanmoy Bhattacharya
and Michael Pennington
for numerous useful discussions; Michael Pennington,
in addition, for pointing
out the existence of the transcription errors in the $K\pi$ parameters
of Ref.~\cite{LASS}; and W. Dunwoodie for confirming the
revised values of these parameters.
The ongoing support of the Natural Sciences and
Engineering Research Council of Canada, and the hospitality and
support of the
Special Research Centre for the Subatomic Structure of Matter at the
University of Adelaide, where much of this work was completed, are also
gratefully acknowledged.

\end{document}